\documentclass[prl,superscriptaddress,nofootinbib,preprint]{revtex4}

\usepackage{amsmath}
\usepackage{blindtext}
\usepackage{algorithm}
\usepackage{algorithmic}
\usepackage{amssymb}
\usepackage{graphicx}
\usepackage{color}
\usepackage{array}
\usepackage[makeroom]{cancel}
\usepackage{bbm}
\usepackage{changes}
\usepackage{epstopdf}

\def\ket#1{\left| #1 \right\rangle}

\newcommand{\beq}{\begin{equation}}
\newcommand{\eeq}{\end{equation}}

\begin{document}
\title{Experimental preparation and verification of quantum money}
\author{Jian-Yu Guan}
\affiliation{Shanghai Branch, Hefei National Laboratory for Physical Sciences at Microscale and Department of Modern Physics, University of Science and Technology of China, Hefei, Anhui 230026, P. R. China}
\affiliation{CAS Center for Excellence and Synergetic Innovation Center in Quantum Information and Quantum Physics, Shanghai Branch, University of Science and Technology of China, Hefei, Anhui 230026, China}
\author{Juan Miguel Arrazola}
\affiliation{Centre for Quantum Technologies, National University of Singapore, 3 Science Drive 2, Singapore 117543}
\author{Ryan Amiri}
\affiliation{SUPA, Institute of Photonics and Quantum Sciences, Heriot-Watt University, Edinburgh EH14 4AS, United Kingdom}
\author{Weijun Zhang}
\author{Hao Li}
\author{Lixing You}
\author{Zhen Wang}
\affiliation{State Key Laboratory of Functional Materials for Informatics, Shanghai Institute of Microsystem and Information Technology, Chinese Academy of Sciences, Shanghai 200050, P.~R.~China}
\author{Qiang Zhang}
\affiliation{Shanghai Branch, Hefei National Laboratory for Physical Sciences at Microscale and Department of Modern Physics, University of Science and Technology of China, Hefei, Anhui 230026, P. R. China}
\affiliation{CAS Center for Excellence and Synergetic Innovation Center in Quantum Information and Quantum Physics, Shanghai Branch, University of Science and Technology of China, Hefei, Anhui 230026, China}
\author{Jian-Wei Pan}
\affiliation{Shanghai Branch, Hefei National Laboratory for Physical Sciences at Microscale and Department of Modern Physics, University of Science and Technology of China, Hefei, Anhui 230026, P. R. China}
\affiliation{CAS Center for Excellence and Synergetic Innovation Center in Quantum Information and Quantum Physics, Shanghai Branch, University of Science and Technology of China, Hefei, Anhui 230026, China}

\begin{abstract}
A quantum money scheme enables a trusted bank to provide untrusted users with verifiable quantum banknotes that cannot be forged. In this work, we report an experimental demonstration of the preparation and verification of unforgeable quantum banknotes. We employ a security analysis that takes experimental imperfections fully into account. We measure a total of $3.6\times10^6$ states in one verification round, limiting the forging probability to $10^{-7}$ based on the security analysis. Our results demonstrate the feasibility of preparing and verifying quantum banknotes using currently available experimental techniques.
\end{abstract}

\maketitle

Remarkable progress has been made in quantum cryptography since its inception several decades ago. Quantum key distribution is widely considered to be one of the first practical quantum technologies \cite{BB84,scarani2009security,diamanti2016practical}, while many other protocols are beginning to shift from theoretical proposals to experimental demonstrations. Examples of these are developments in quantum signature schemes \cite{AWKA2016,AWA2015,amiri2015unconditionally,collins2017experimental,collins2016experimental,yin2017experimental}, quantum fingerprinting \cite{arrazolaqfp,xu2015experimental,GX16}, secure quantum computation \cite{broadbent2009universal,barz2012demonstration,barz2013experimental,greganti2016demonstration}, covert communication \cite{sanguinetti2016perfectly,bash2015quantum,arrazola2016covert,bradler2016absolutely}, and bit commitment \cite{ng2012experimental,lunghi2013exp,liu2014experimental,verbanis201624}. Despite these advances, quantum money -- the first quantum cryptography protocol to be proposed -- has only recently started to enter the realm of possible experimental implementations.

Quantum money was first introduced in a seminal paper by Wiesner in 1970 \cite{W1983}. The goal of any quantum money scheme is to enable a trusted authority, the bank, to provide untrusted users with verifiable banknotes that cannot be forged. Many variants of Wiesner's original scheme were found to be vulnerable to so-called ``adaptive attacks" \cite{A09,L2010,Brodutch2016}, which motivated the formulation of new quantum money protocols which are provably secure against unbounded quantum adversaries. Similarly, progress was made in developing simpler protocols that take into account experimental limitations. In Ref. \cite{Gav2012}, a secure quantum money protocol was proposed requiring only classical communication between a verifier and the bank. The issue of tolerance to experimental errors was first addressed in Ref. \cite{PYJ2011}, with further developments in Refs. \cite{GK2015}. Recently, a practical protocol with nearly optimal noise tolerance was proposed in Ref. \cite{amiri2017quantum}. These developments have lead to the first quantum money experiments, with a demonstration of forging in Wiesner's original scheme \cite{bartkiewicz2017experimental}. An unforgeable demonstration of quantum money remains experimentally challenging.

In this work, we present an experimental implementation of the quantum money scheme of Ref. \cite{amiri2017quantum}, demonstrating the entire life-cycle of the quantum states contained in a quantum banknote: from preparation using a laser source and phase modulation, to verification using passive linear optics. We perform a security analysis of the protocol that takes full account of experimental imperfections. The setup allows for fast and efficient verification of quantum banknotes, compatible with on-chip realizations and storage in quantum memories, which may be performed in the future.

In the remainder of this paper, we give a detailed description of the quantum money protocol, including the bank's algorithm for preparing the quantum banknotes and the verification procedure of the holders. We then describe the experimental setup for state preparation and verification, and finally give the results of calibration of the protocol as well as the verification of the banknotes.

\textit{Quantum money protocol.--} Any scheme for producing unforgeable quantum banknotes consists of a procedure from the bank to prepare the banknotes and a method to verify their authenticity. In this work, we implement the practical quantum money scheme of Ref. \cite{amiri2017quantum}, which is based on hidden matching quantum retrieval games (QRGs) \cite{Gav2012,JM}. In these QRGs, the bank encodes a four-bit classical string $x=x_1x_2x_3x_4$ into a sequence of coherent states with amplitude $\alpha$ of the form
\beq\label{Eq:States}
\ket{\alpha,x}:=\ket{(-1)^{x_1}\alpha}\ket{(-1)^{x_2}\alpha}\ket{(-1)^{x_3}\alpha}\ket{(-1)^{x_4}\alpha}.
\eeq
The verifier's goal is to perform a measurement on $\ket{\alpha,x}$ that allows her to retrieve the value of the parity bit $b=x_i \oplus x_j$, where the possible $(i,j)$ pairs are specified by the matchings $M_1 = \{(1,2), (3,4) \}$, $M_2 = \{(1,3), (2,4) \}$, and $M_3 = \{(1,4), (2,3)\}$. This measurement can be done by employing unbalanced Mach-Zehnder interferometers, as explained in detail later in this paper. These hidden matching QRGs form the building block of the quantum money protocol, as described below.

\textbf{Banknote preparation}

\noindent 1. The bank independently and randomly chooses $N$ strings of four bits which we will call $x^{1}, ..., x^{N}$.

\noindent 2. The bank creates $N$ quantum states $\ket{\alpha,x^1}$, $\ket{\alpha,x^2}$, $\ldots,\ket{\alpha,x^N}$, which constitute the quantum banknote. The bank assigns a unique serial number to the banknote for identification.

\noindent 3. The bank creates a classical binary register $r$ and initializes it to $0^N$. This register keeps a record of the states that have been previously used in the verification.

\noindent 4. The bank creates a counter variable $s$ and initializes it to $0$. This counter keeps a record of the number of verification attempts for the banknote. The bank also have a pre-defined maximum number of allowed verifications $T$. The banknote should be returned to bank if $s\ge T$.

\textbf{Banknote verification}

Before the verification step, the protocol need to be calibrated to give the total efficiency $\eta$ and the base error rate $\beta$ of the measurement setup. The detail of calibration can be seen in Supplemental Materials. The holder must give enough correct outcomes otherwise the verifier aborts the protocol. The detailed steps are described below.

1. The holder randomly chooses a subset of indices $L\subset [N]$ of size $l=|L|$ such that $r_k =0$ for each $k \in L$. For each $k\in L$, the holder sets the corresponding bit of $r$ to $1$, indicates these states will be measured.

2. For each $k\in L$, the holder picks a matching $M^k$ at random from $M_1,M_2,M_3$ and applies the corresponding measurement to obtain outcome $b_k=x_i\oplus x_j$. If there is no outcome, we set $b_i = \emptyset$. The number of successful outcomes is defined as $l^\prime$.

3. The bank sets the efficiency threshold to be $\eta - \epsilon$, where $\epsilon > 0$ is a small positive security parameter. If $l^\prime < l_{min} := (\eta - \epsilon)l$,  the verifier aborts the protocol.

4. The holder sends all triplets $(k, (i,j), b_k)$ to the bank, who checks that $s<T$.

5. For each $k$, the bank checks whether the answer is correct by comparing $(k, (i,j), b_k)$ to the secret $x^k$ values. The bank sets an error threshold to be $\beta+\delta$, where $\delta$ is a small positive constant. The bank accepts the banknote as valid only if fewer than $l^\prime(\beta+\delta)$ of the answers are incorrect.

6. The bank updates $s$ to $s+1$.\\

\begin{figure}[t!]
\begin{center}
\includegraphics[width=0.75\columnwidth]{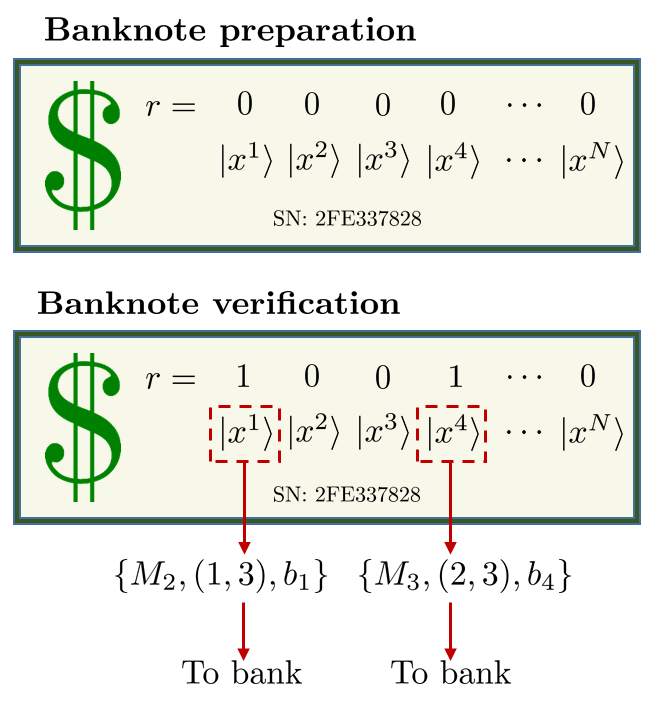}
\caption{(Color online) Schematic illustration of the quantum money protocol. The bank produces $N$ quantum states $\ket{x^1},\ket{x^2},\ldots,\ket{x^N}$ according to a random secret string $x=x^1x^2\cdots x^N$. The bank also assigns a unique serial number to the banknote and creates a register $r$ that records whether each state has been used previously for verification. To verify the banknote, a holder randomly selects $l$ quantum states. For each state, the holder randomly selects one of the three matchings $M_1, M_2, M_3$ and performs the corresponding measurement. The outcomes consist of a matching pair $(i,j)$ and a parity bit $b$, which are recorded and sent for comparison with bank's secret string $x$. The banknote is accepted as valid if the error rate observed by the bank is sufficiently low.}
\label{Fig:Protocol}
\end{center}
\end{figure}

\begin{figure*}[t]
\includegraphics[width=0.92\textwidth]{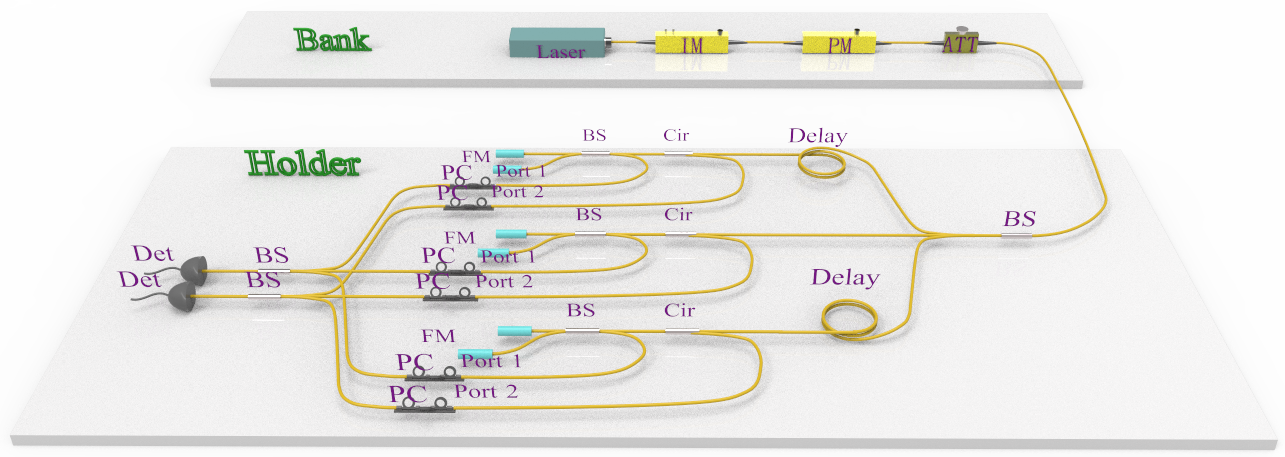}
\caption{Experimental setup for generating and verifying quantum banknotes. A laser source produces sequences of coherent states which are modulated in phase according to a secret string $x$ and attenuated to an amplitude of $|\alpha|^2=0.25$. The signals are passively split into three arms using a $1\times 3$ beam-splitter and routed to three Mach-Zehnder interferometers. The interferometers -- which respectively have delays of 2~ns, 4~ns, and 6~ns -- are implemented using a single beam-splitter and Faraday mirrors. The delays are depicted in the figure in terms of the varying length of the lower arm of the interferometer connected to the mirror. Delays of lengths 5~m, 10~m are placed to distinguish the outputs of each interferometer, which are recombined using a $1\times 3$ beam-splitter and measured using superconducting nanowire single-photon detectors. IM: Intensity Modulator. PM: Phase Modulator. ATT: Attenuator. BS: beam-splitter ($2\times 2$ and $1\times 3$). Cir: Circulator. FM: Faraday Mirror. PC: Polariser Controller.}\label{fig:1}
\centering
\end{figure*}

The complete protocol is illustrated in Fig. \ref{Fig:Protocol}. The security and correctness of this protocol was proven in Ref. \cite{amiri2017quantum}, where it was shown that an honest verifier will correctly verify a valid banknote except with probability
\beq\label{Eq:Ver}
\texttt{P}(\text{Ver fails}) \leq \exp \left[-2l_{min}\delta^2\right] + \exp[-2l\epsilon^2]
\eeq
and the probability that an adversary can forge a banknote is bounded by
\beq\label{Eq:Forge}
\texttt{P}(\text{Forge}) \leq \exp [-2\frac{\epsilon^2}{\eta^2}l] + \exp [-2l\epsilon^2] + \exp [-2l_{min}\delta^2].
\eeq
Both of these probabilities decrease exponentially in the protocol parameter $l$. Besides, in our security analysis, the forging probability is larger than the probability of failing the verification, so we only take the former into account.

The parameters $\epsilon$ and $\delta$ are chosen to minimize the value of $l$ necessary to achieve a given security level. We use $e_{\text{min}}$ to represent the minimum average verification error rate for a forged coin. A natural choice for $\delta$ is $\delta = (e_{\text{min}}-\beta)/2$, i.e. half of the gap between the average error rate for a genuine coin and a forged coin. Security can always be obtained as long as $\beta<e_{\text{min}}$. In the protocol, $e_{\text{min}}$ is bounded by \cite{amiri2017quantum}
\beq \label{eq:emin}
e_{\text{min}} \geq \left(\frac{\frac{1}{6} - \frac{3\epsilon}{2\eta}}{1-\frac{3\epsilon}{\eta}}\right)\frac{4|\alpha|^2 e^{-4|\alpha|^2}}{1-e^{-4|\alpha|^2}}.
\eeq
The optimal choice of $\epsilon$ depends on the system parameters and is calculated numerically. In what follows, we outline the experimental procedure to implement the quantum money protocol.

\textit{Experimental implementation.---} To prepare the banknotes, we employ a continuous-wave (CW) laser with a wavelength of 1550.12~nm and a linewidth of 50~kHz. The laser is modulated to generate a block of four continuous pulses using an intensity modulator.  Every block is 96~ns long while each individual pulse has a width of 2~ns so that the blocks of four pulses occupy a total of 8~ns. This resulting low duty ratio is chosen to allow for time multiplexing while still allowing a large repetition rate of 10 MHz. The block's length is much shorter than laser's coherent time, so all the pulses in a block have the same global phase. The phase information, which depends on the bank's secret key $x$, is encoded on each pulse via a phase modulator to create the states as in Eq. \eqref{Eq:States}. The secret key $x$ is generated using a quantum random number generator and stored in a pulse pattern generator (PPG) with amplifiers. The key data is then used to modulate the phase of the pulses. Finally, an attenuator adjusts the average photon number to an optimal level of $|\alpha|^2=0.25$. Each block is now a quantum state of the banknote, which is transmitted to the holder.

For verification of the banknote, the holder randomly selects a subset of all pulses and measures them. In this proof-of-principle experiment, this is done by measuring all states and selecting a random subset of all outcomes. Verification requires the holder to choose randomly between three different measurements, each corresponding to a different matching. This is achieved using a $1\times 3$ beam-splitter (BS) to passively select between three Mach-Zehnder interferometers with delays of 2~ns, 4~ns, and 6~ns. The interferometers employ Faraday mirrors and a single beam-splitter to combine all possible pairings in the matchings. The result of interference implies phase relation between the corresponding two pulses. Thus, this allows the holder to retrieve information about the parity of the secret bits encoded in their phase.

Since the pulses in each block are separated by 2~ns, the 2~ns interferometer performs interference of the pairs $(1,2), (2,3), (3,4)$, the 4~ns interferes pairs $(1,3), (2,4)$, and the last interferometer interferes the pairing $(1,4)$. This covers all six pairs in the matchings, allowing the holder to perform the banknote verification. At the output of the beam-splitter, delays of lengths 0~m, 5~m and 10~m are introduced to distinguish the outputs of the interferometers by their arrival time. Two $1\times 3$ beam-splitters are used to recombine the output light of the interferometers. We use two superconducting nanowire single-photon detectors (SNSPDs) for detection. The SNSPDs have a desired polarization which corresponds to its maximum detection efficiency of $70\%$. At each output port of the interferometers, we use a polarization controller to adjust the polarization. Finally, the detection events are recorded by a time-digital converter (TDC) for analysis. The experimental setup is shown in Fig.~\ref{fig:1}.

\begin{figure}
  \centering
  \includegraphics[width=0.9\columnwidth]{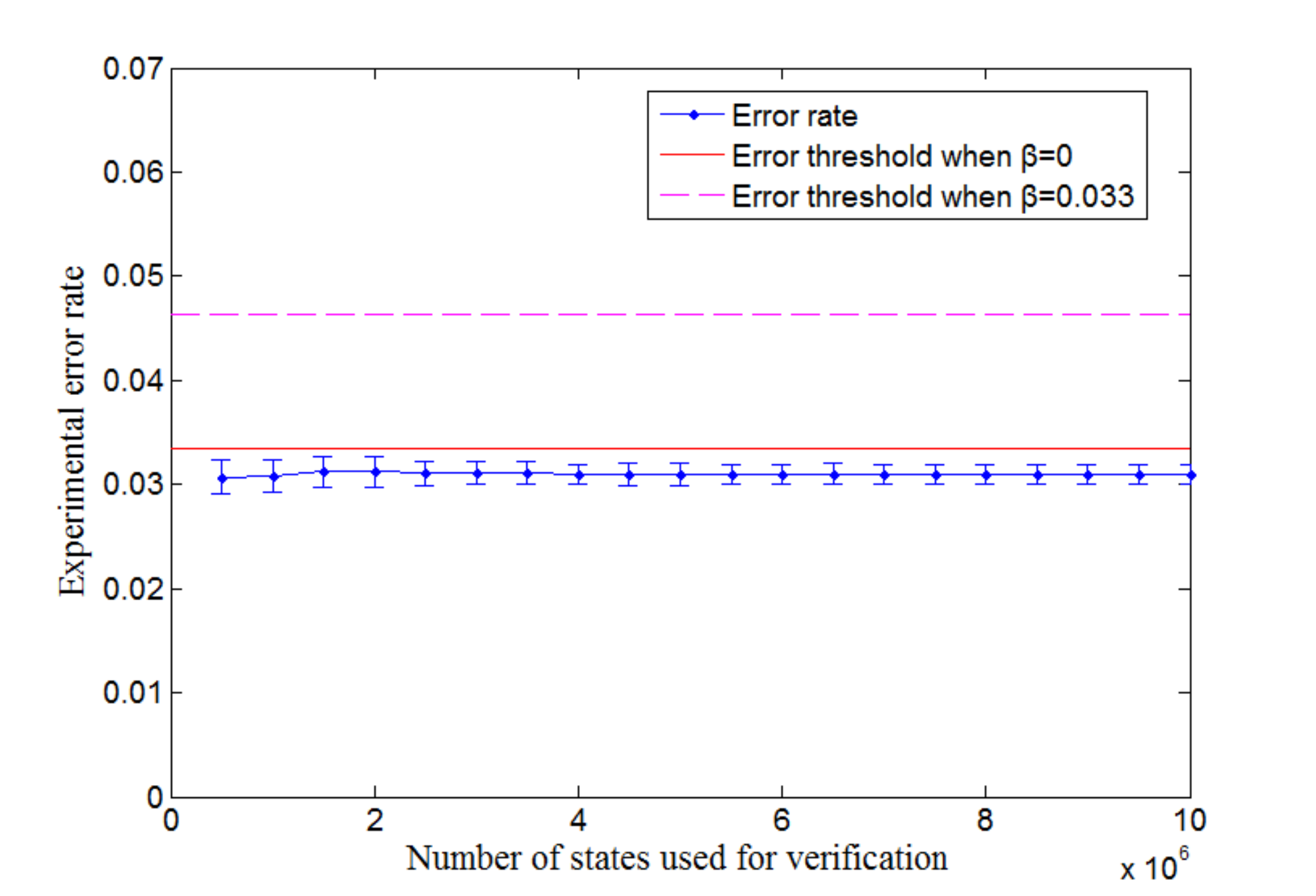}\\
  \caption{Experimental error rate for different values of the number of state used in verification $l$. The standard error is calculated from 10 rounds of experiments. The red line shows the maximum allowed error rate for passing the verification when $\beta=0$, and the magenta dashed line shows the threshold when $\beta=0.033$. The banknotes generated in the experiment pass the verification in both cases.}\label{Fig:Result1}
\end{figure}

\textit{Experimental results.---} As seen in Eqs. \eqref{Eq:Ver} and \eqref{Eq:Forge}, the security levels of the protocol depend partly on two parameters: the overall efficiency $\eta$ and the expected error rate $\beta$. The parameter $\epsilon$ is optimized to achieve the lowest number of measurement states $l$. The detailed method for optimization can be refer to Supplemental Materials. In the calibration step, we obtain $\eta=3.36\%$ and $\beta=0.033$ for our setup, and the optimal value of $\epsilon$ is 0.0018. In a more strict analysis, in our protocol, we can set $\beta=0$, which means that all measurement errors are assumed to come from the adversary attempting to forge the banknote, as opposed to experimental imperfections. The corresponding $\epsilon$ is set to 0.0015. The total efficiency of all experimental rounds lie from $0.0336$ to $0.0339$, and the experimental error rate at different block size is shown in Fig.~\ref{Fig:Result1}. Thus all our experiments can pass the verification.

\begin{figure}
  \centering
  \includegraphics[width=0.9\columnwidth]{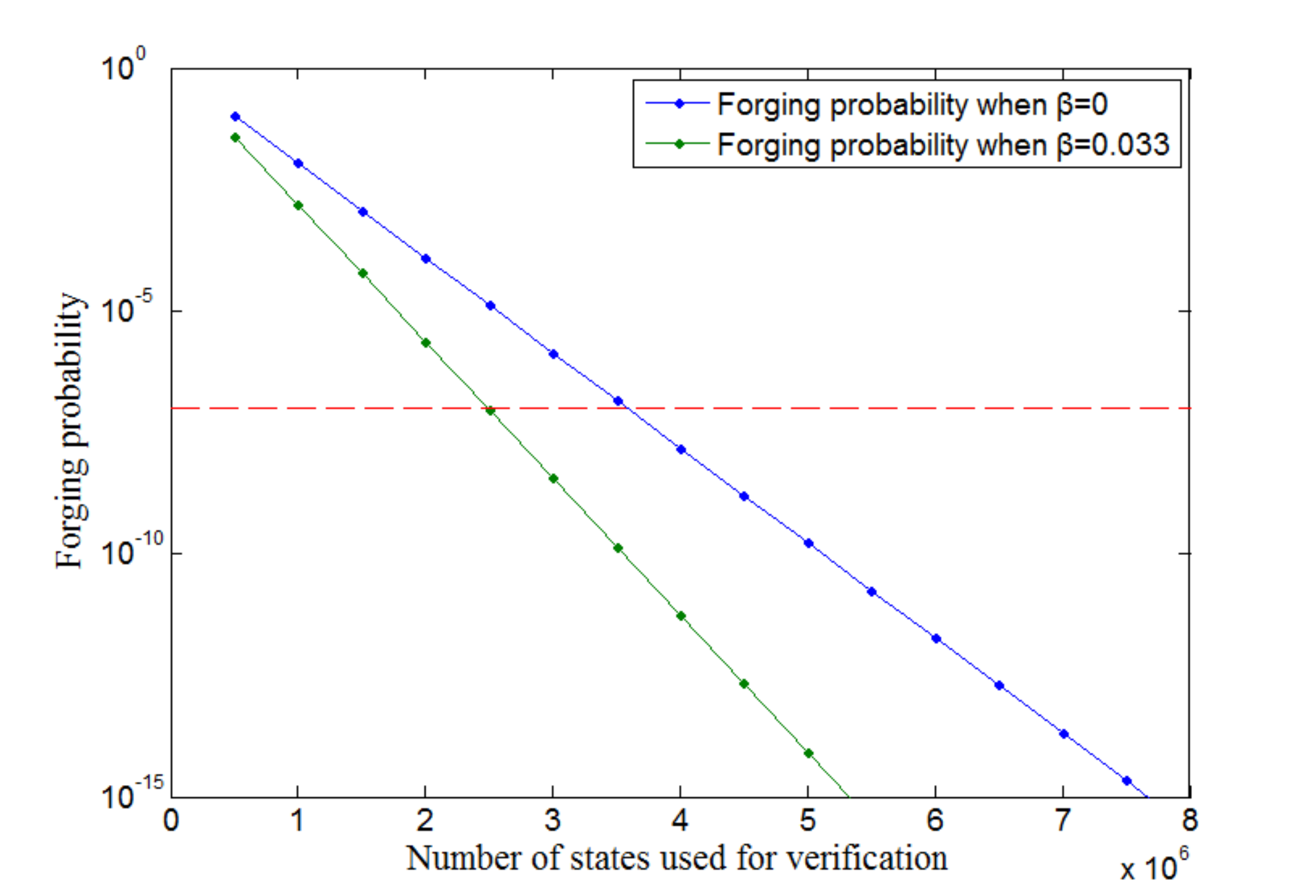}\\
  \caption{Forging probability for different values of the number of state sused in verification $l$. The blue line corresponds to parameters $\beta=0.033$ while the green line corresponds to parameters $\beta=0$. The dashed line represents the $10^{-7}$ target security level.}\label{Fig:Result2}
\end{figure}

The security of the quantum money protocol is quantified by the forging probability, which we set to $10^{-7}$.  The forging probability of our protocol is shown in Fig.~\ref{Fig:Result2}, for the cases $\beta=0.033$ and $\beta=0$. The largest number of states needed occurs for $\beta=0$, where at most $l=3.6\times10^6$ states need to be measured in one verification round to ensure the desired security level, which actually decreases exponentially with $l$. This takes less than 350~ms in our experiment.

\textit{Discussion.---} We have reported an experimental implementation of the preparation and verification of unforgeable quantum banknotes. As a proof-of-principle demonstration, our results show that these ingredients of quantum money protocols are technologically viable. In order to reach full applicability of quantum money schemes, it is crucial to be able to store the quantum states constituting the banknote in quantum memories. This remains a daunting challenge, but progress has been made rapidly in developing memories capable of storing the quantum states of optical modes as required by this money protocol \cite{sinclair2014spectral,gundougan2015solid,lvovsky2009optical}. Additionally, the interferometers used for verification are suitable for an implementation using integrated optics, which would allow a convenient method method for verifying quantum banknotes. Beyond their application to quantum money, our results demonstrate an implementation of quantum retrieval games (QRGs), which have the potential to be used as building blocks in other cryptographic protocols. This is an area worth exploring further. For example, it is intriguing to note the similarity between hidden matching QRGs and round-robin differential phase-shift QKD \cite{guan2015experimental,li2016experimental,sasaki2014practical}, a connection that may lead to new insights into these protocols.

Note added: we became aware of a relevant work \cite{bozzio2017experimental} when preparing the manuscript. Their work is based on the theoretical proposal of Ref. \cite{PYJ2011} using polarization qubits while we utilize high dimensional time-bin qudits based on the protocol of Ref. \cite{amiri2017quantum}.

\bibliography{QuantumMoneyBib}
\bibliographystyle{apsrev}

\end{document}